\documentclass[useAMS,usenatbib]{mn2e}
\usepackage{graphicx}

\title[Effect of the Solar motion on flux of LP-comets]{The effect of the
Solar motion on the flux of long-period comets} \author[E. Gardner, P.
Nurmi, C. Flynn and S. Mikkola]{E. Gardner$^1$\thanks{E-mail:
esgard@utu.fi}, P. Nurmi$^1$, C. Flynn$^2$ and S. Mikkola$^1$\\
$^1$Tuorla Observatory, Department of Physics and Astronomy,
University of Turku, V\"ais\"al\"antie 20, FI-21500, Piikki\"o,
Finland\\ $^2$Finnish Centre for Astronomy with ESO (FINCA),
University of Turku, V\"ais\"al\"antie 20, FI-21500, Piikki\"o,
Finland }
    
\begin{document}

\date{Accepted 2010 September 16.  Received 2010 September 16; in
original form 2010 June 24}

\pagerange{\pageref{firstpage}--\pageref{lastpage}} \pubyear{2010}

\maketitle

\label{firstpage}

\begin{abstract}
The long-term dynamics of Oort cloud comets are studied under the
influence of both the radial and the vertical components of the
Galactic tidal field. Sporadic dynamical perturbation processes are
ignored, such as passing stars, since we aim to study the influence of
just the axisymmetric Galactic tidal field on the cometary motion and
how it changes in time. We use a model of the Galaxy with a disc,
bulge and dark halo, and a local disc density, and disc scale length
constrained to fit the best available observational constraints. By
integrating a few million of cometary orbits over 1 Gyr, we calculate
the time variable flux of Oort cloud comets that enter the inner Solar
System, for the cases of a constant Galactic tidal field, and a
realistically varying tidal field which is a function of the Sun's
orbit. The applied method calculates the evolution of the
comets by using first-order averaged mean elements. We find that the
periodicity in the cometary flux is complicated and quasi-periodic.
The amplitude of the variations in the flux are of order 30 \%. The
radial motion of the Sun is the chief cause of this behaviour, and
should be taken into account when the Galactic influence on the Oort
cloud comets is studied.

\end{abstract}


\section{Introduction} 

Strong observational evidence supports the idea that the inner Solar
system is subject to a steady flux of `new' comets which originate
from the `Oort cloud' \citep{Oort50}. The semi-major axes of these
comets are thought to evolve under the influence of external forces
such as the Galactic tidal field and passing stars. The comets evolve
from a typical $a_{orig} \simeq 3 \times 10^4$ AU to either hyperbolic
orbits or to larger binding energies, depending on the orbital
evolution during the cometary encounters with the Solar System
planets. During one orbital evolution, comets typically experience
perturbations that change the comet's orbit into a short-period or
hyperbolic orbit, which leads to a subsequent ejection into
interstellar space \citep{Nurmi2001}.


The Galactic potential exerts a force on Oort cloud comets, and is important
for the steady state flux of comets with $a > 20000$ AU,
e.g. \cite{Byl1983,Heisler86,Matese89}.  The Galactic tidal force has a
dominant component that is perpendicular to the Galactic plane; the radial
component is 10 times weaker than the perpendicular component
\cite{Heisler86}. For this reason, many studies
(e.g. \citealt{Matese1995,Wick2008}) have assumed that the radial Galactic
tidal field component is negligible. Other studies have included the radial
component, for more accurate modelling of cometary motion, such as in
\cite{Matese96,Brasser01}. The radial component of the tide has been found to
have an effect on the long-term evolution of the comets' perihelia, on the
distribution of the longitudes of the perihelion \citep{Matese96}, and the
origin of chaos in the cometary motion \citep{Breiter2008}. In recent large
scale simulations, \cite{Rickman08} found that a fundamental role is played by
perturbations due to passing stars on comets, contrary to the investigations
during the previous two decades, starting with \cite{Heisler86}. The stellar
perturbations, do of course act together with the Galactic tide. 

The topic of this paper is the effect on comets due to the Galactic tide
alone. We use a realistic model of the local Galaxy, which is well constrained
by observations, which contains a disc, bulge and halo. We follow the orbit of
the Sun in this model using recent constraints on the Solar motion. This motion
allows us to compute the change in the vertical and radial components of the
Galactic tide with time, and the change in the tidal force due to the radial
motion of the Sun is fully accounted for.

Qualitatively, the tidal effect on the cometary orbits can be
evaluated by studying the change in angular momentum averaged over one
orbit. The Galactic tidal force periodically changes the angular
momentum ($J=\sqrt{GM_\odot a (1-e^2)}$) of the Oort cloud comets
\citep{Fernandez91}. The angular momentum of the comets changes as
\citep{Heisler86}:

\begin{equation}\label{DJ}
\frac{dJ}{dt}=-\frac{5\pi \rho_0}{G
\mathrm{M}_{\odot}^{2}}L^2(L^2-J^2)[1-(J_{z}^{2}/J^2)]\sin2\omega_g ~.
\end{equation}

Here, $\omega_g$ is the Galactic argument of perihelion, $J_{z}$ is
the $z$-component of angular momentum, and is perpendicular to the
Galactic plane, $L = \sqrt{\mu a}$, $\rho_0$ is the local mass
density, and $G$ is the gravitational constant. Note that this
equation is valid for first order mean elements under the action of
axially symmetric disc tides. The periodically
changing angular momentum causes variations mainly in perihelion
distance $q$, for comets in near-parabolic orbits, since $q \approx
J^2/(2G\mathrm{M}_{\odot})$. The typical assumptions in the cometary
flux calculations due to the Galactic tidal force suppose that the
local tidal field is axisymmetric, perpendicular to the mid-plane, and
adiabatically changing \citep{Matese1992}. 

The aim of many of these studies is to correlate the motion of the Sun in the
Galaxy with phenomena on the Earth, such as mass extinctions of species, the
cratering record and climate change. An extensive review of this topic has been
made recently by \cite{Bailer-Jones09}.

For example, early studies have shown that the ages of well dated
impact craters on Earth are not distributed randomly, but that there
is a possible 28 Myr \citep{Alvarez1984} or 30 $\pm$ 1 Myr periodicity
in crater ages over the past 250 Myr \citep{Rampino1984}. Since then,
several authors have claimed that there is a significant periodic
signal present, but the periods differ quite a lot from study to
study. The signal is the most prominent for 40 large, well-dated
craters that are up to 250 Myr old \citep{Napier2006}, but the period
is difficult to measure, with estimates of between 24-26 Myr
\citep{Napier2006}, 30 Myr \citep{Napier2006,Stothers2006}, 36 Myr
\citep{Napier2006,Stothers2006}, 38 Myr \citep{Yabushita2004,Wick2008}
and 42 Myr \citep{Napier2006}.

In some studies, the reliability of the signal is questioned altogether, based
on the inaccuracy of the age estimates of the impact craters, possible biases
caused by rounding the ages of craters, and the small number of craters
\citep{Grieve1996,Jetsu2000}.

By critically reviewing many studies that have tried to connect the Solar
motion and periodicity in terrestrial phenomena such as biodiversity, impact
cratering and climate change, \cite{Bailer-Jones09} has concluded that there is
little evidence to support these connections. By studying the artificial
cratering data \cite{Lyytinen2009} came to the same conclusion, that the
reliable detection of any periodicity is currently impossible with the
existing cratering data.

In this study, we statistically analyse how the Galactic tidal force
changes cometary orbits over 1 Gyr, using numerical simulations. The 1
Gyr time-scale is long enough to observe changes in the Galactic tide
due to both the radial and vertical motion of the Sun. A simple
axisymmetric Galactic potential is adopted. To our knowledge, there
has been no study to date, in which the effect of both radial and
vertical components, in a time varying Galactic potential (via the
variation in mass density $\rho$), has been analysed in detail. Our
purpose is to study the statistical effects of the complete Galactic
potential to the Oort cloud comets in detail, especially concentrating
on the comets that enter the Solar System ($q<30$ AU). In particular,
we analyse the differences in cometary motion for when the tidal field
is constant, and when it varies as the Sun moves in a realistic orbit
in a fairly realistic Galactic potential.  We find that it is
important to include the radial motion of the Sun in the calculations,
since the local density varies significantly as the Sun moves towards
and away from the Galactic centre.

\section{Methods} 

The method of simulation requires two, traditionally separate,
components. The first is to simulate the motion of the Sun around the
Galaxy. The second involves the evolution of the orbits of comets in
the Oort cloud. We integrate the motion of the Sun in an axisymmetric
Galactic potential. The method of integration of the comets is
described in \cite{Mikkolanurmi}. The method calculates the evolution
of the comets by using first-order averaged mean elements.  We do not
include the random perturbations caused by the planets, since the aim
is to identify the tidal effects of the Galactic potential on the flux
of comets reaching the inner Solar System.

\subsection{The Galactic potential} 

The Galactic potential consists of a disc, bulge, and dark halo, and
is partially described in \cite{Gardner10}. The model
used here differs most notably from \cite{Gardner10} in the treatment
of the disc, as well as a slightly modified dark halo. We
noticed that the vertical density profile of the disc, from
\cite{Gardner10}, does not accurately reproduce the observational
profile from \cite{Holmberg04}. We proceeded to modify our disc-model
by changing a few of the model's parameters, and adding three more
Miyamoto-Nagai potentials to the model, to emulate a very thin layer
of gas in the disc. The equation for the full potential is:
\begin{equation}
\Phi = \Phi_H + \Phi_C + \Phi_D + \Phi_g~,
\end{equation}
\begin{equation}
 \Phi_H = \frac{1}{2}V_h^2 {\rm ln}(r^2 + r_0^2)~ ,
\end{equation}
\begin{equation}
 \Phi_C = - \frac{GM_{C_1}}{ \sqrt{r^2+r^2_{C_1}} } -
 \frac{GM_{C_2}}{\sqrt{r^2+r^2_{C_2}}}~ ,~ \\
\end{equation}
\begin{equation}
 \Phi_D = \sum_{i=1}^3~~\frac{-GM_{d_i}}{\sqrt{(R^2+(a_{d_i}+\sqrt{(z^2+b^2)})^2)}}~ ,\\
\end{equation}
\begin{equation}
 \Phi_g = \sum_{n=1}^3~~\frac{-GM_{g_n}}{\sqrt{R^2+[a_{d_n}+\sqrt{(z^2+b_g^2)}]^2}} ~,
\end{equation}

where $G$ is the gravitational constant, $R$ is the
Galactocentric radius, and $z$ is the height. The modified parameters
are in Table \ref{newparameters}. The vertical density profile of the
model can be seen in Fig. \ref{heightprofile}.

\begin{table}
\caption{Parameters of the full potential.}\label{newparameters}
\begin{center}
\begin{tabular}{lcr}
\hline
Property & value & Unit\\
\hline
$V_h$ & 220 & km s$^{-1}$\\
$r_0$ & 10 & kpc \\
$b$ & 0.45  & kpc \\
$b_g$ & 0.12 & kpc \\
$r_{C_1}$ & 2.7 & kpc\\
$r_{C_2}$ & 0.42 & kpc\\
$a_{d_1}$ & 5.81 & kpc\\
$a_{d_2}$ & 17.43 & kpc\\
$a_{d_3}$ & 34.86 & kpc\\
$M_{C_1}$ & 3 & 10$^9$ M$_\odot$\\
$M_{C_2}$ & 16 & 10$^{9}$ M$_\odot$\\      
$M_{d_1}$ & 66.06$\times 10^9$  & M$_\odot$\\
$M_{d_2}$ &$-$59.05 $\times 10^9$  & M$_\odot$\\
$M_{d_3}$ & 22.97$\times 10^9$  & M$_\odot$\\
$M_{g_1}$ & 18.63$\times 10^9$  & M$_\odot$\\
$M_{g_2}$ & $-$16.66$\times 10^9$  & M$_\odot$\\
$M_{g_3}$ & 6.48$\times 10^9$  & M$_\odot$\\
\hline
\end{tabular}
\end{center}
\end{table}

The mass density at the Sun is
$0.11$ M$_\odot/$pc$^3$, consistent with observational constraints
(0.10 $\pm$ 0.01 M$_\odot/$pc$^3$ \citet{Holmberg00} and 0.105 $\pm$
0.005 M$_\odot/$pc$^3$ \cite{Korchagin03}). This corresponds to a
nominal T$_z$=$83\times10^6$ year period for small amplitude simple
harmonic motion in the vertical direction. The surface density of disc
matter in the model is 54.9 M$_\odot/$pc$^2$, compared with a measured
disc surface density of $56 \pm 6 $M$_\odot/$pc$^2$ \cite{Holmberg04}.
The adopted current position of the Sun in the model, $(R,z)_\odot$, 
is (8,0) kpc. The local circular velocity of the model is 221 km
s$^{-1}$.

\subsubsection{Density and rotation curve}

\begin{figure}
\includegraphics[width=84mm]{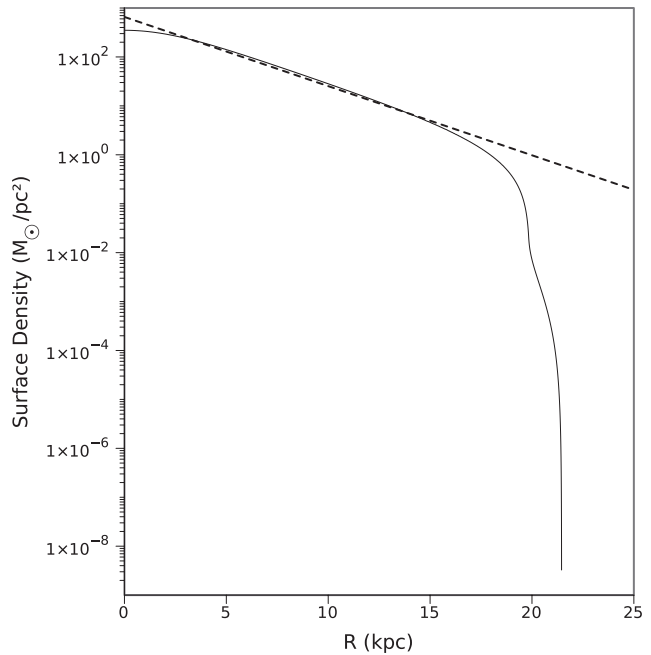}
\caption{The surface density of the disc component as a function of
  Galactocentric radius. The dashed line corresponds to an exponential density
  falloff of 3 kpc, which is a good fit to the model over a wide range of
  radii. Note that the density truncates strongly at 18
  kpc.}\label{discprofile}
\end{figure}

\begin{figure}
\includegraphics[width=\columnwidth]{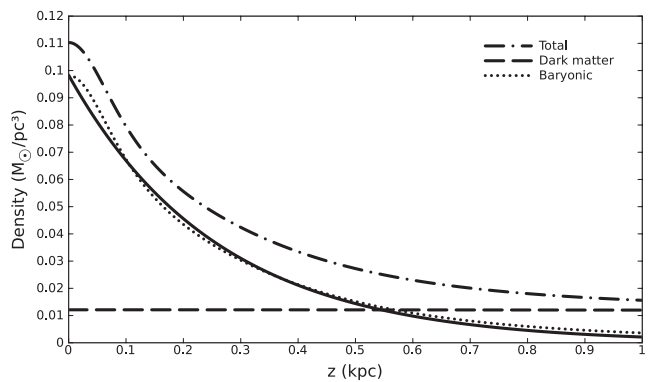}
\caption{Vertical density of the model, at the Sun ($R$ = 8 kpc).
The dotted line represents the baryonic contribution in the model, the
dashed line the dark matter contribution, and the dashed-dotted line
the total density of the model at a certain height ($z$). The solid
line corresponds to an exponential fit to the baryonic component of
the model of 0.25 kpc.}\label{heightprofile}
\end{figure}

\begin{figure}
\includegraphics[height=84mm,angle=270]{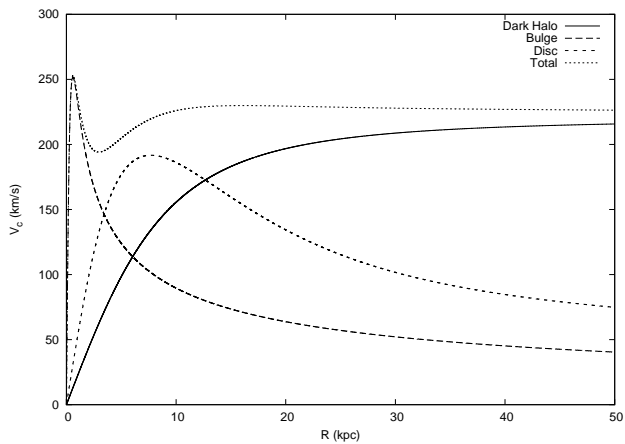}
\caption{Rotation curve for the model Milky Way (dotted), and the different
  contributions of the disc (dashed), bulge (long dashed), and dark halo
  (solid).}\label{vcirc}
\end{figure}

Fig. \ref{discprofile} shows the surface density of the disc with
radius ($R$), the change in density with height ($z$) at the Sun ($R=
8 kpc$) (Fig.
\ref{heightprofile}), and the rotation curve (Fig. \ref{vcirc}), as
these are the two factors that have the most impact on the orbits of
the comets. The disc has a scale-length of 3 kpc, and a local
scale-height of 0.24 kpc, consistent with recent measurements by
\cite{Juric08} using the Sloan Digital Sky Survey (SDSS).

The potential is axisymmetric, so we do not model effects such as
molecular clouds, spiral arms, and bubbles in the interstellar matter
and passing stars. Our interest here is on the global Galactic effects
of the Solar motion on the comets.

\subsubsection{Motion of the Sun}

\begin{figure}
\includegraphics[height=84mm,angle=270]{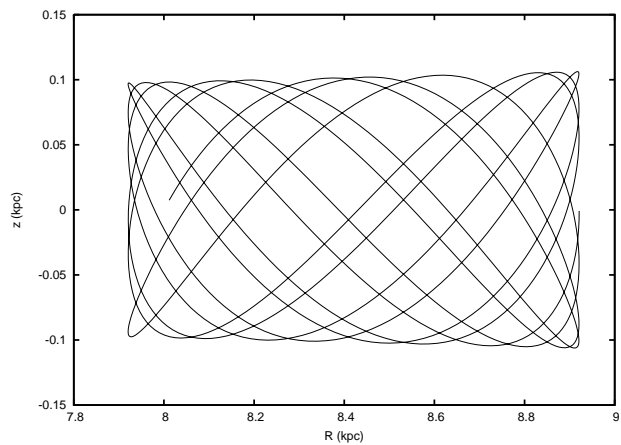}
\caption{Motion of the Sun in the model potential for 1 Gyr, in radial $R$ and
  vertical $z$ components. The initial position of the Sun is $(R,z)_0
  = (8,0)$ kpc.}\label{orbit}
\end{figure}

The orbit of the Sun, using the Solar motion measured by
  \cite{Schonrich10}, where ($U, V, W$)=(11.1, 12.24, 7.25) km
  s$^{-1}$, and assuming ($R,z$)$_\odot$ to be (8,0) kpc, is shown in
  the ($R,z$)-plane in Fig. \ref{orbit}. $U$ is the velocity
  towards the Galactic centre, $V$ is the velocity along rotational
  direction, and $W$ is the velocity perpendicular to the Galactic
  plane. Properties of this orbit, such as eccentricity, radial and
  vertical period and maximum $z$ height, are shown in Table
  \ref{sunmotion}. The eccentricity of the Solar orbit ($e$) is
  defined as:
  ${(R_{\mathrm{max}}-R_{\mathrm{min}})}/{(R_{\mathrm{max}}+R_{\mathrm{min}})}$.

\begin{table}
\caption{Eccentricity, $e$, maximum vertical height,
  $z_{\mathrm{max}}$, radial oscillation period, $T_R$, and vertical
  oscillation period, $T_z$, of the Sun in the adopted potential. The
  adopted solar motion is that of
  \protect\cite{Schonrich10}.}\label{sunmotion}
\begin{center}
\begin{tabular}{lcr}
\hline
Property & value & Unit\\
\hline
$e$ & 0.059 $\pm$ 0.003 & \\
$z_{\mathrm{max}}$ & 0.102 $\pm$ 0.006 & kpc\\
$T_R$ & 149 $\pm$ 1 & Myr\\
$T_z$ & 85 $\pm$ 4 & Myr\\
\hline
\end{tabular}
\end{center}
\end{table}

\subsection{Variation of the tidal parameters caused by the motion of
the Sun}\label{motion-section}

Our integration method for cometary motion is based on the computation of
secular motion of the comet in quadratic perturbation by the Galactic
potential. The method used is presented in \citet{Mikkolanurmi}. As the comet
orbits the Sun under the Galactic potential it experiences a force
$\bmath{F}$ per unit
mass:

\begin{equation}
\bmath{F}=-\frac{G
  M_{\odot}}{r^3}\bmath{r}-G_1x\bmath{x}-G_2y\bmath{y}-G_3z\bmath{z},
\end{equation}

where $G_1=-(A-B)(3A+B)$, $G_2=(A-B)^2$, and $G_3=4 \pi G \rho(R,z)
-2(B^2-A^2)$. Here $r$ is the Sun-Comet distance, $\rho(R,z)$ is the local mass
density, and $G$ is the gravitational constant \citep{Heisler86}. $A$ and $B$
are the Oort constants, and are obtained from the Galactic model. Usually the
radial components ($\bmath{x},\bmath{y})$ are neglected, so that $G_1 = G_2 =
0$.

We will examine the effect on comets in two particular cases. Firstly
we study what we call the 'dynamic' case, in which the Galactic tide
changes realistically along the Solar orbit, for the case that the Sun
oscillates both vertically and radially in the potential.  Secondly,
we assume that there is a 'constant' tidal field (i.e. the
tidal field does not change as the Sun moves around the Galaxy).  In
the constant case the Sun moves on a flat, circular orbit.

\begin{figure}
\includegraphics[angle=-90,width=84mm]{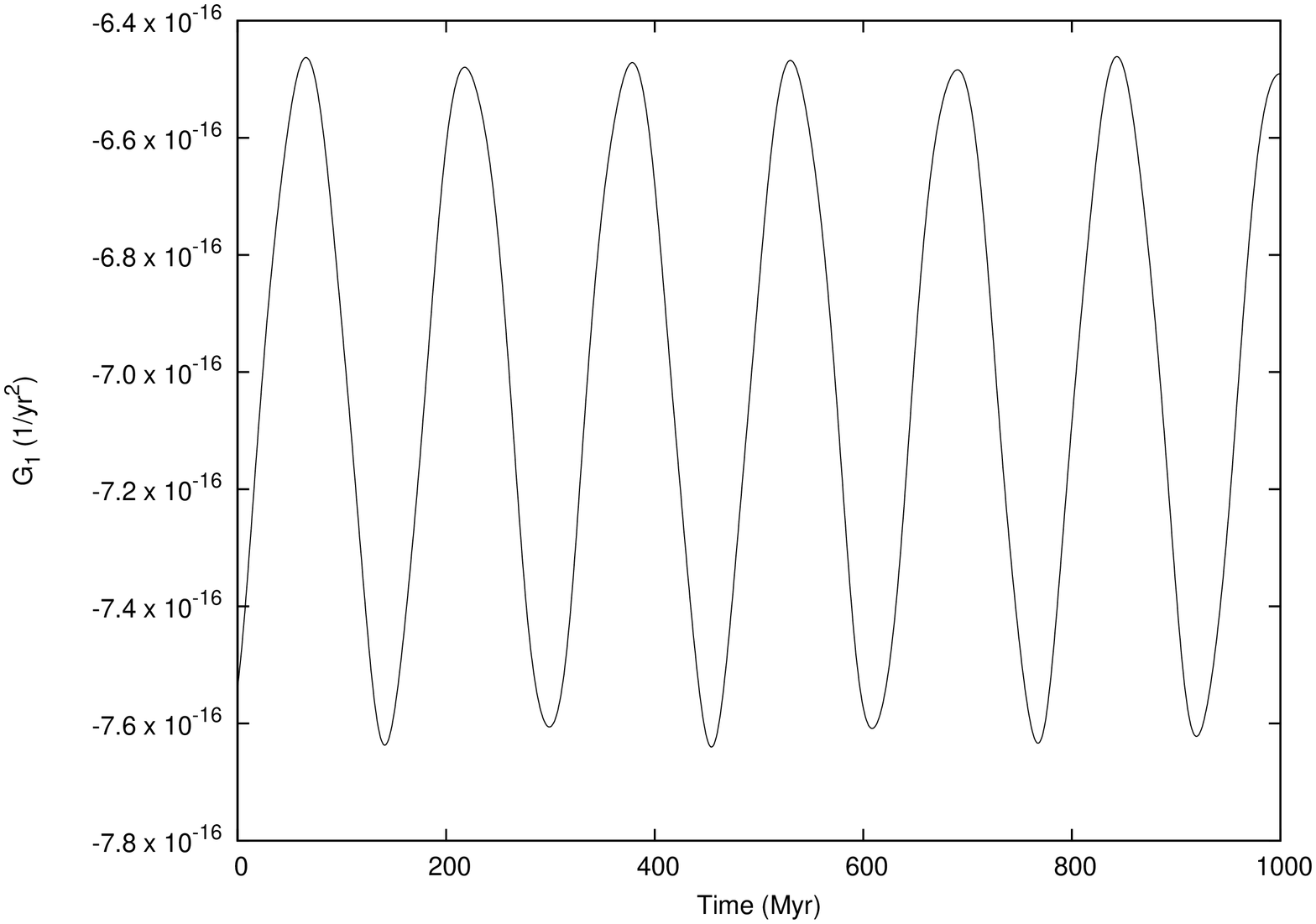}\\
\includegraphics[angle=-90,width=84mm]{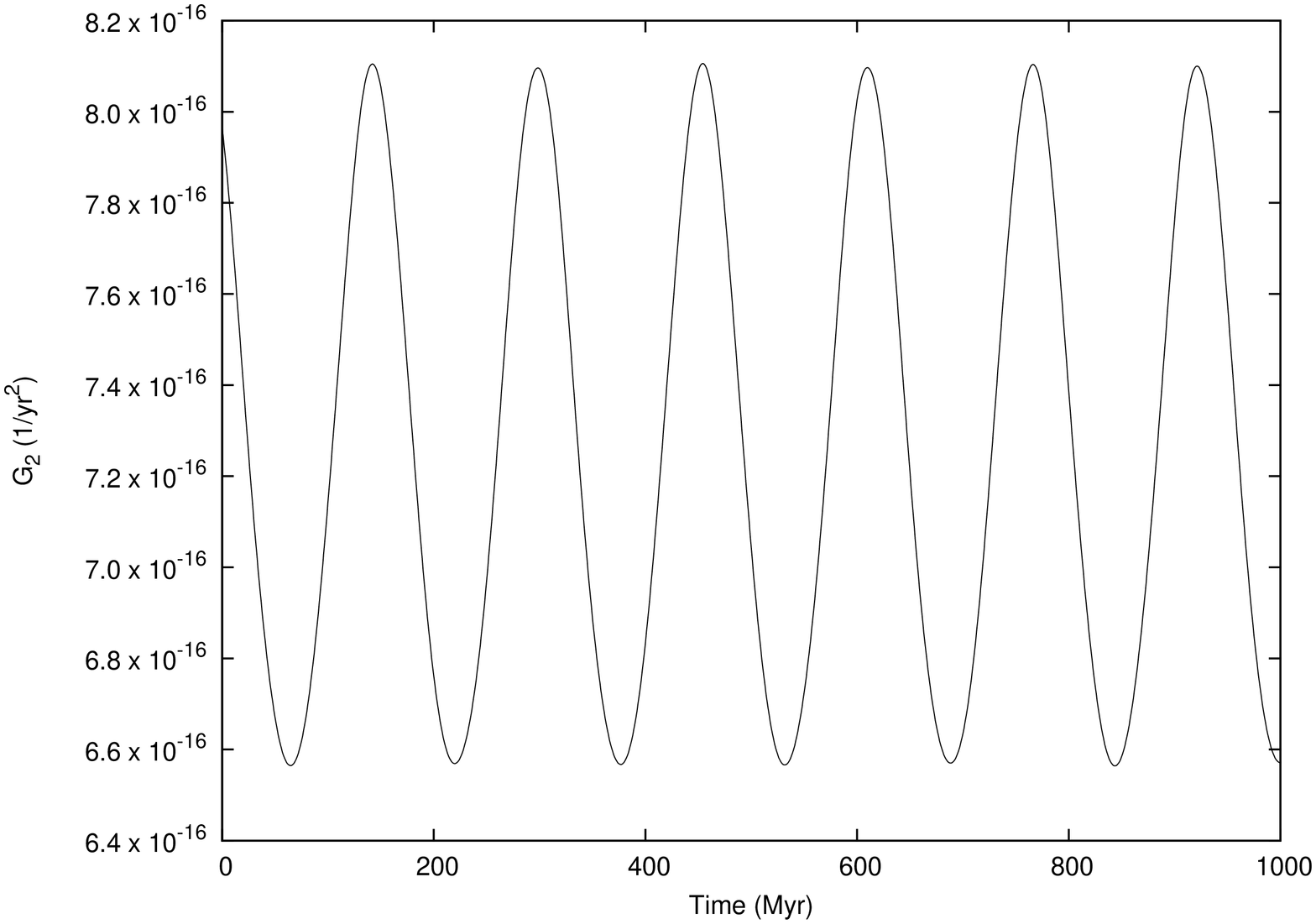}\\
\includegraphics[angle=-90,width=84mm]{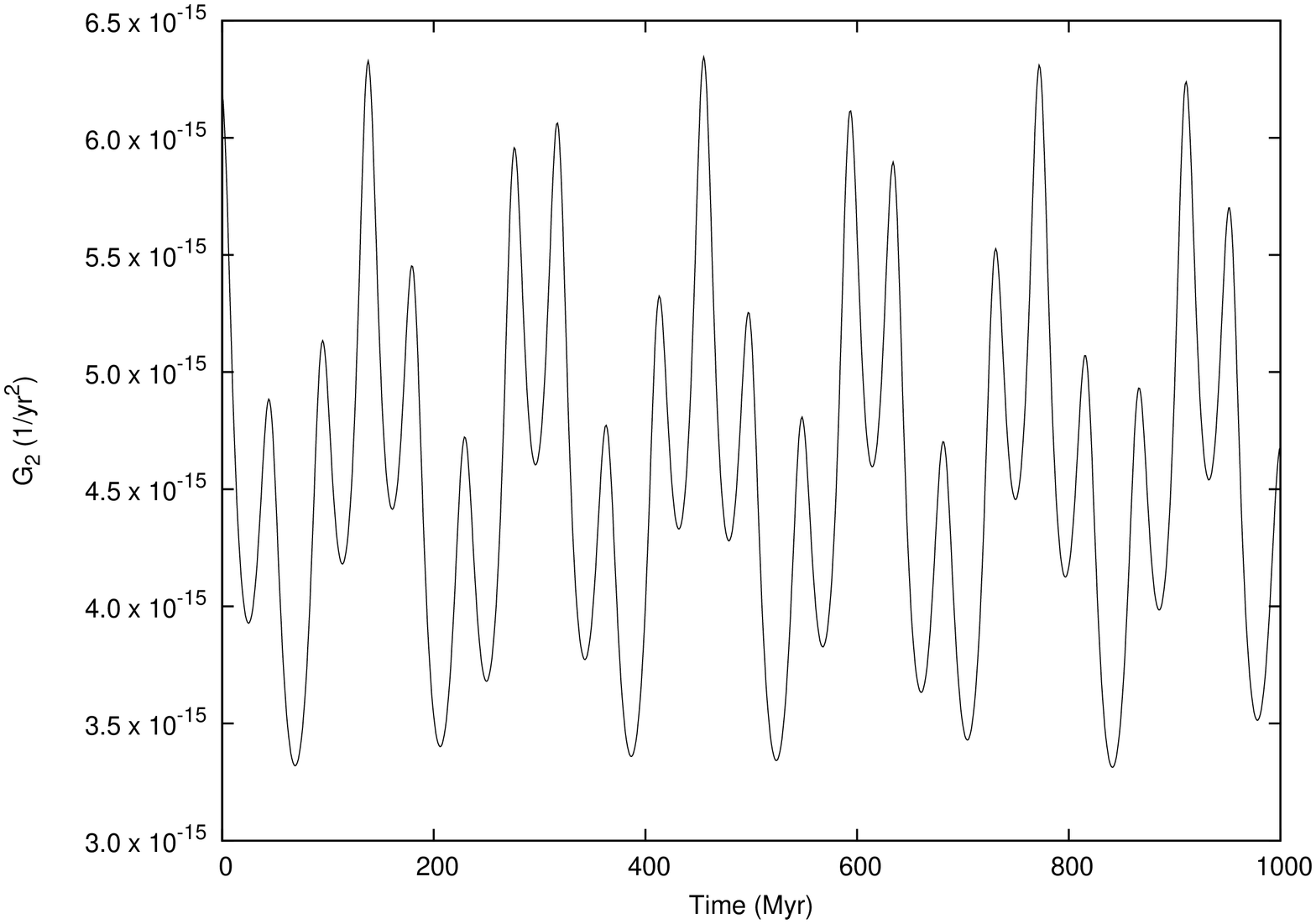}
\caption{Variation of the parameters $G_1$ (top panel), $G_2$ (centre panel),
  and $G_3$ (bottom panel) along the Solar orbit in the 'dynamic' case
  (i.e. including full vertical and radial motions) over 1 Gyr.}\label{gs}
\end{figure}

We integrate the Solar orbit in the 'dynamic' case for 1 Gyr, sampling
the values of the $G$-parameters every 100 kyr: these are shown in
Fig. \ref{gs}. The changes in $G_3$ are dominated by the changes in
local density during the orbit, since the changes in $A$ and $B$ over
the orbit are not very large. Fig. \ref{gs} (top and centre panel)
shows how the changing values of the Oort constants, $A$ and $B$,
affect the values of $G_1$ and $G_2$.

\begin{figure}
\includegraphics[height=84mm, angle=270]{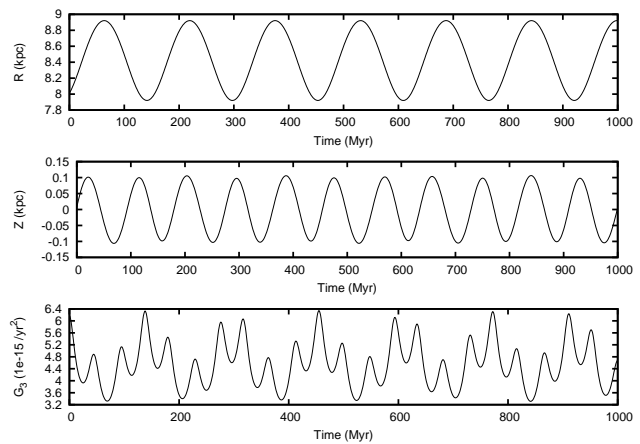}
\caption{Galactocentric radius $R$, vertical height $z$ and $G_3$ for the
  'dynamic' case of the Solar motion over 1 Gyr in the Galactic adopted
  model. Upper panel: the radial motion. Middle panel: vertical motion. Lower
  panel: $G_3$.}\label{dynamic-changes}
\end{figure}

Fig. \ref{dynamic-changes} shows the combined effects of the radial
($R$) and vertical motion ($z$) of the orbit, on $G_3$.  It is clear
that $G_3$ increases in two situations: when the radial position is
the closest to the Galactic centre, and when the vertical motion
crosses the mid-plane.  Due to the slightly eccentric motion of the
orbit, it is the radial component of the motion which dominates the
changes in local density ($\rho(R,z)$), rather than the vertical
motion. This is partly because we adopt the Solar motions of
\cite{Schonrich10}, which produces a mildly eccentric orbit for the
Sun ($e=0.059\pm0.003$): it oscillates between Galactocentric radii of
7.9 and 8.9 kpc. As such, the evolution of $G_3$ depends on the
radial motion and the vertical motion, both being of equal magnitude (Fig.
\ref{dynamic-changes}). 


In the second case studied, the Sun is set on a perfectly circular and flat
orbit, so that the local mass density does not change with time.  For the
'constant' case, the values of $G_i$ are the mean values from the 'dynamical'
case, and have the values $G_1 = -7.00897 \times 10^{-16}$, $G_2 =
7.27613 \times 10^{-16}$, and $G_3 = 4.55749 \times 10^{-15} \mathrm{yr}^{-2}$.

\section{Simulations of the comets}

Observations support the idea that the \textit{in situ} flux of new
comets is roughly linear with respect to heliocentric distance, up to
the distance of Jupiter \citep{Hughes2001}. Comets with highly
eccentric orbits with randomised directions of motion have a
$q$-distribution that is flat \citep{Opik66}. If a long period comet
has a perihelion distance between $\sim$ 10$-$15 AU, it is quickly
ejected to interstellar space or perturbed so that it becomes a
short-period comet \citep{Wiegert99}. From the perspective of the Oort
cloud, the comets are removed from the Oort cloud and from the loss
cone to the orbital parameter distribution \citep{Fernandez81}. The
loss cone (lc) is the population of comets that have orbits that will allow
them to penetrate the planetary system, making it possible to observe
them \citep{Hills81}. Planetary perturbations move comets efficiently
from $q$ values less than $q_\mathrm{lc} \simeq 15$ AU into either hyperbolic
orbits or into orbits that are more tightly bound to the solar system
\citep{Hills81}. In the steady state situation, the new comets are
distributed uniformly to the perihelion distances of $q \leq q_\mathrm{lc}$,
for $a>30000$ AU, while $q>q_\mathrm{lc}$ comets come also from the inner
region. For this reason, in all of our simulations we have assumed
that initial perihelion distances are distributed uniformly outside
the loss cone. The inclination distribution of the outer Oort cloud of
comets is isotropic (uniform in $\cos i$).  All the other angular
elements are uniformly distributed between $[0-2\pi]$.
\subsection{Structure of the Oort cloud}
An important issue in studying the structure of the Oort cloud, is
what energy distribution to adopt for the comets. We assume that the
density of comets between 3000 AU and 50000 AU in the Oort cloud is
proportional to $1/r^{3.5 \pm 0.5}$, so that the number of comets $N$
is $dN \propto 1/r^{\alpha} dr$, where $\alpha = 1.5 \pm 0.5$
\citep{Duncan87}. The conclusions of this paper are not particularly
sensitive to the adopted value of $\alpha$. The existence of an inner
Oort cloud has been speculated upon in many studies although there is
no direct evidence for it \citep{Hills81}. An inner Oort cloud is the
extension of the Kuiper belt, filling the gap between the Kuiper belt
and the outer Oort cloud. Semi-major axes in the inner Oort cloud are
typically between 50$-$15000 AU \citep{Leto2008}. The steady state
flux from the inner Oort cloud cannot be uniformly distributed in
perihelion distance, since the planetary perturbations move comets
efficiently from $q$ values less than $q_{lc} \simeq 15$ AU to either
hyperbolic orbits or into orbits that are more tightly bound to the
solar system \citep{Hills81}. In the steady state situation, the new
comets come uniformly to perihelion distances $q \leq q_{lc}$ when
$a>30000$ AU, while the $q>q_{lc}$ comets also come from the inner
region.

\subsection{Simulation parameters of the comets}
Due to this complicated picture, we study different semi-major axes in
separate simulations, and evaluate the efficiency of tidal injection
in each simulation. We chose initial sample conditions for the Oort
cloud comets, setting the semi-major axis to be 10000, 20000, 30000,
40000, 50000, and 60000 AU. We choose the comet's eccentricity ($e$) randomly,
so that the resulting values of $q$ would be uniformly distributed
from 35 to the value of the semi-major axis ($a$). All the other
parameters, $\cos(i)$, $\Omega$, $\omega$, and the initial eccentric
anomaly were all chosen randomly in appropriate intervals. The number
of comets in each simulation is $10^6$ at each of the sampled
semi-major axes. For the computation of the numbers of comets reaching
the inner Solar System, the results from each semi-major axis are
normalised to the adopted number density law. 

\subsection{Simulations of the Galactic tide}
To study the effect of the Galactic potential on cometary motion, we
chose two hypothetical solar systems: a `constant' background density,
where the Sun would be on a pure circular orbit with no vertical
motion, and a realistic 'dynamic' Solar orbit. In all systems, we
analyse the flux of Oort cloud comets into the Solar System. This
means that we consider a comet to have been detected in the inner
Solar System when its $q$ is within 30 AU, and it has a heliocentric
radius of less than 1000 AU. The last criterion is important, as the
osculating elements of comets can evolve to have $q$ $\leq$ 30 AU, far
away from the Sun, and evolve to more than 30 AU, without ever
entering the inner Solar System. Since the minimum of the osculating
$q$ is at the aphelion, the criterion $q \leq$ 30 AU is approximate.
For this reason, we also check if the comet is actually approaching
the aphelion, by choosing $r < 1000$ AU. This requires use of the mean
anomaly, which is approximately calculated from the derived time
co-ordinate. Comets which have been detected are removed from the
simulation. We do not replace them with new comets.  

\section{Results} 

\begin{figure}
\includegraphics[width=84mm]{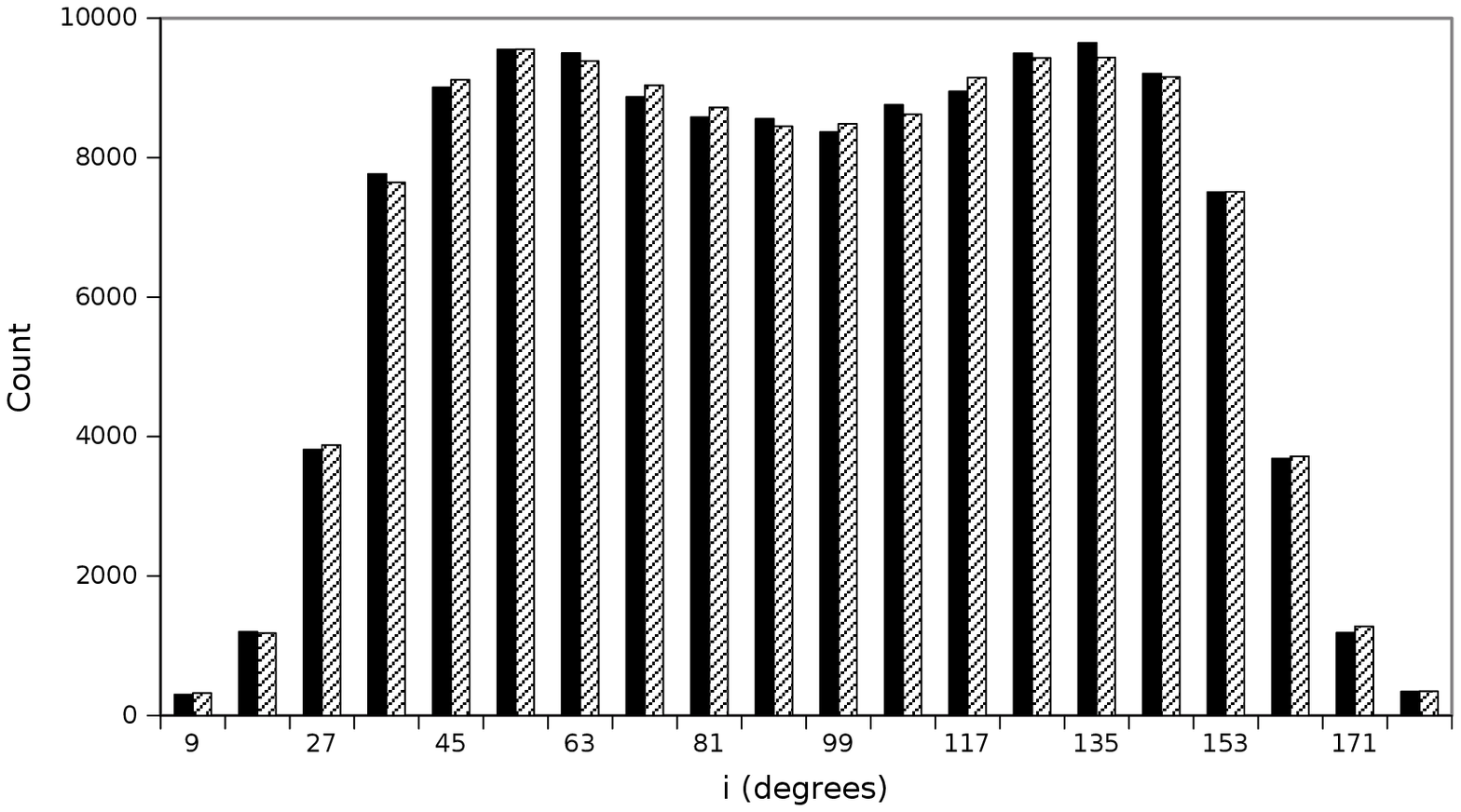}\\
\includegraphics[width=84mm]{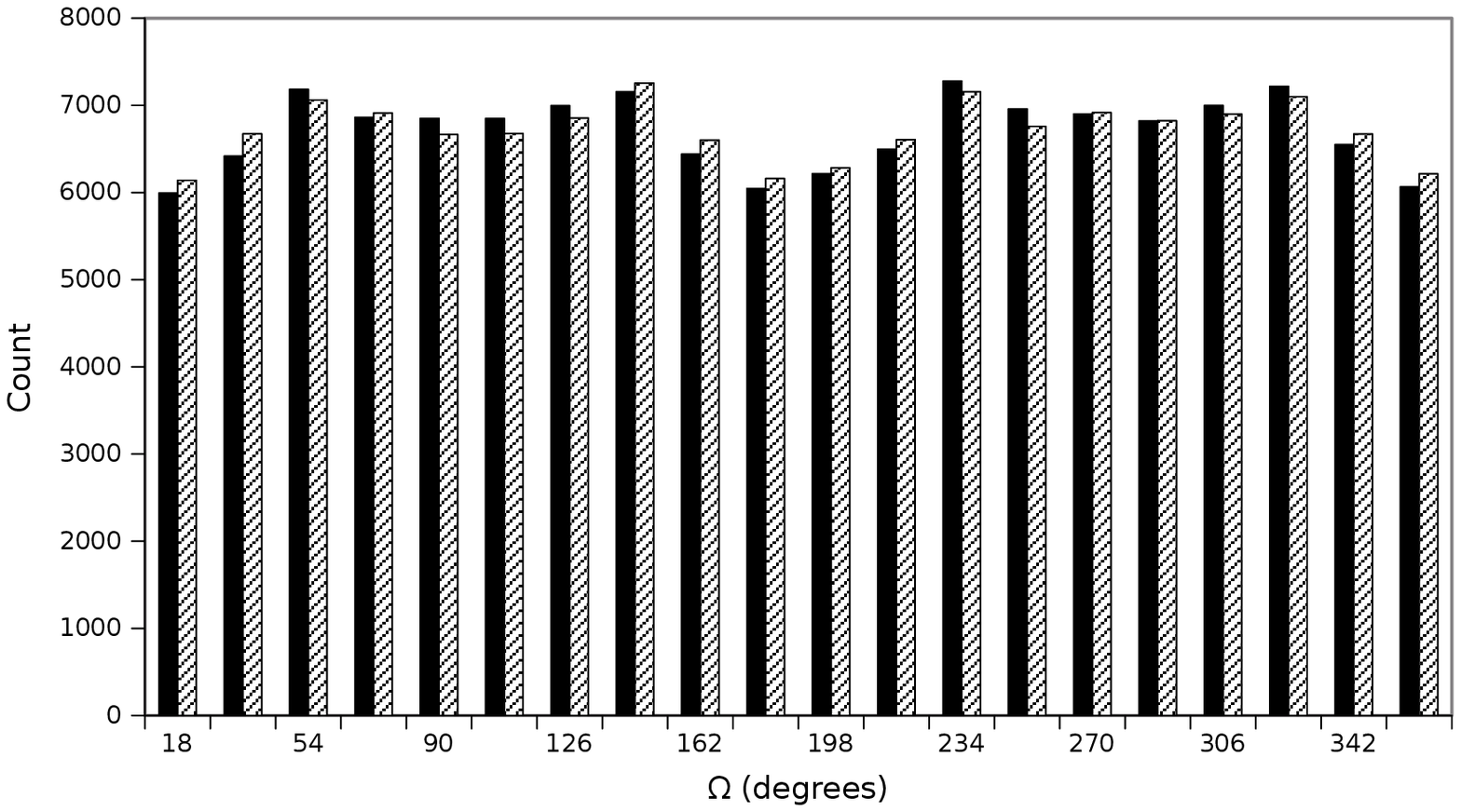}\\
\includegraphics[width=84mm]{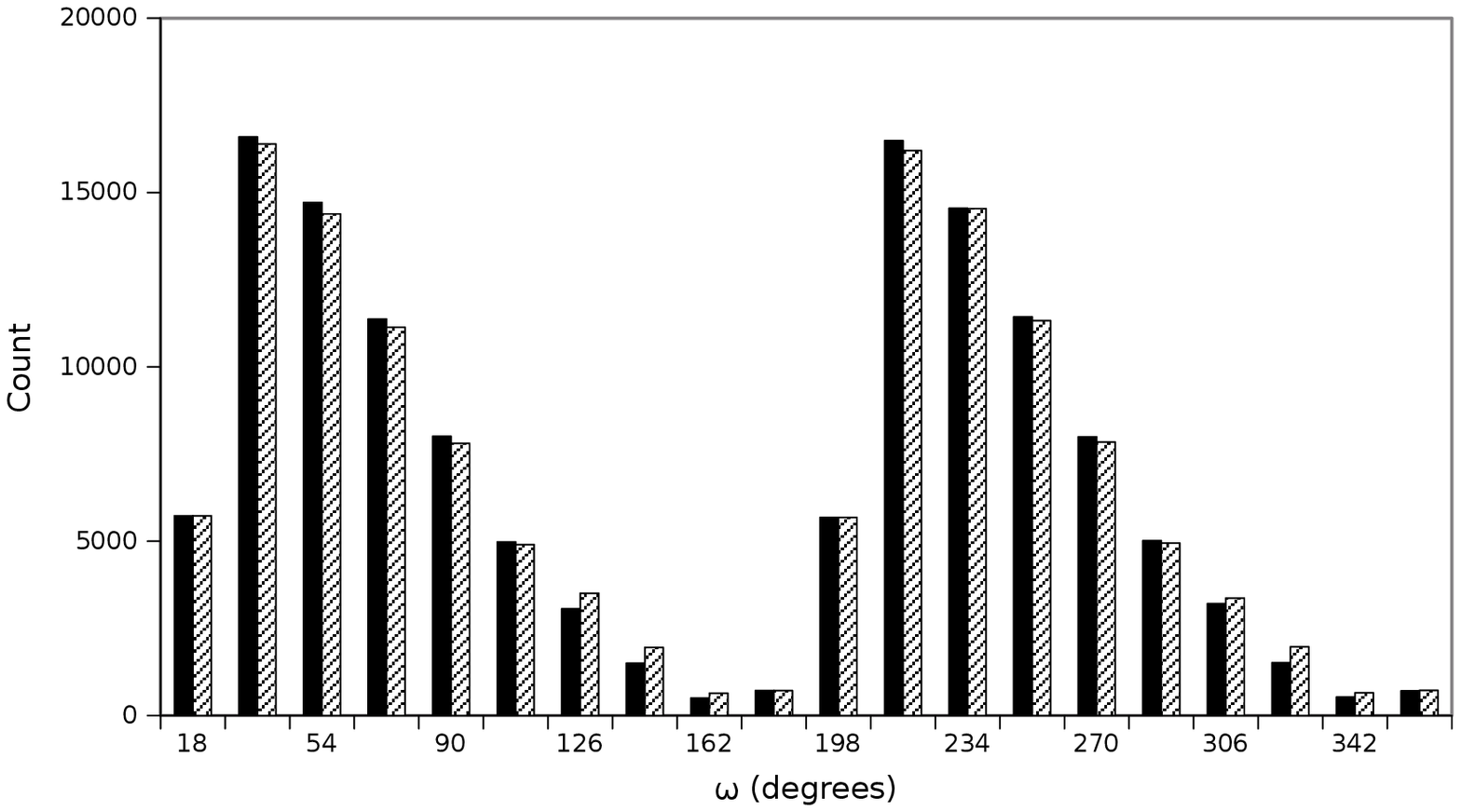}
\caption{Distribution of the orbital elements $i$, $\Omega$, and $\omega$ for
  $a = 30000$. The solid column represents constant case, and the shaded column
  the dynamic case. There is no significant difference between the two
  distributions.}\label{elements}
\end{figure}

\begin{figure}
\includegraphics[width=84mm]{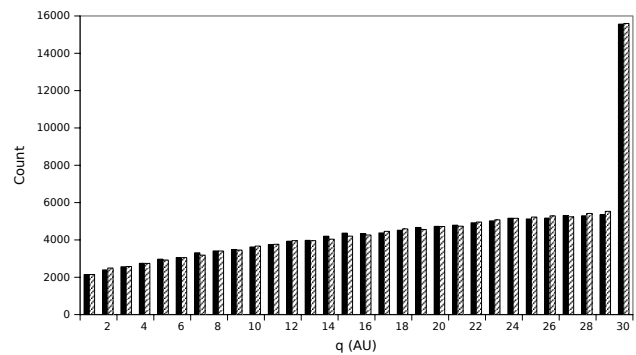} \caption{Distribution of the perihelion
  distance $q$ for the incoming comets, for $a = 30000$. The solid column
  represents the constant case, and the shaded column the dynamic case. There
  is no significant difference between the two distributions.}\label{qs}
\end{figure}

Our aim is to determine the effect of the Galactic tide on Oort cloud
comets by separately adopting a constant local density and a varying
local density. Examining the flux of comets into the inner Solar
System (Table \ref{fluxes}), we find that there is no significant
difference between the two models. Most of the detected comets come
from the middle (30$-$40 kAU) range of semi-major axes. Likewise, the
resulting distribution of orbital elements, for comets coming into the
Solar System, is the same in both the constant and dynamic cases. Fig.
\ref{elements} shows example distributions for the simulation cases
where the semi-major axis is 30000 AU. The gap at $\omega = 0$
and 180 is caused by the $\sin 2\omega_g$ term going to 0 in Equation
\ref{DJ}. Similarly, the factor $1 - (J_{z}^{2}/J^2)$ goes to zero at
$i$=0 or 180, since $J_z=J \cos i$. The $q$-distribution of incoming
comets shows no significant difference between the two cases, as seen
in Fig. \ref{qs}. The high peak at $q=30$ AU, in Fig.
\ref{qs} is caused by the slow evolution of $q$, at $a=30000$ AU.
Finally, the total flux of comets entering the inner-Solar system is
not significantly different between the two cases. For the 'constant'
case, from a source pool of 1.8 million comets in the simulated Oort
cloud, we find that approximately 120 comets per Myr reach the inner
Solar System (which here means comets with $q<30$ AU). Assuming that
there are about $10^{12}$ comets in the Oort cloud \citep{Wiegert99},
this corresponds to about 70 comets per year with $q<30$ AU. This is a
bit higher than the observed number of new comets with $q<30$ AU,
which is about 20 per year, assuming a cometary flux of 0.65 $\pm$
0.18 yr$^{-1}$ AU$^{-1}$ \citep{Fernandez10}.

\begin{table}
\caption{Relative fraction of comets entering the inner Solar System,
from each of the simulated semi-major axis values, for the 'constant'
and 'dynamic' cases.}\label{fluxes}
\begin{center}
\begin{tabular}{lcc}
\hline
Semi-major axis (AU) & constant & dynamic\\
\hline

10000 & 0.0729 & 0.0732\\
20000 & 0.1803 & 0.1804\\
30000 & 0.2405 & 0.2405\\
40000 & 0.2381 & 0.2369\\
50000 & 0.1504 & 0.1507\\
60000 & 0.1179 & 0.1184\\
\hline
\end{tabular}
\end{center}
\end{table}

\subsection{Time evolution of the cometary flux}\label{timeevolution-section}

\begin{figure}
\includegraphics[width=84mm]{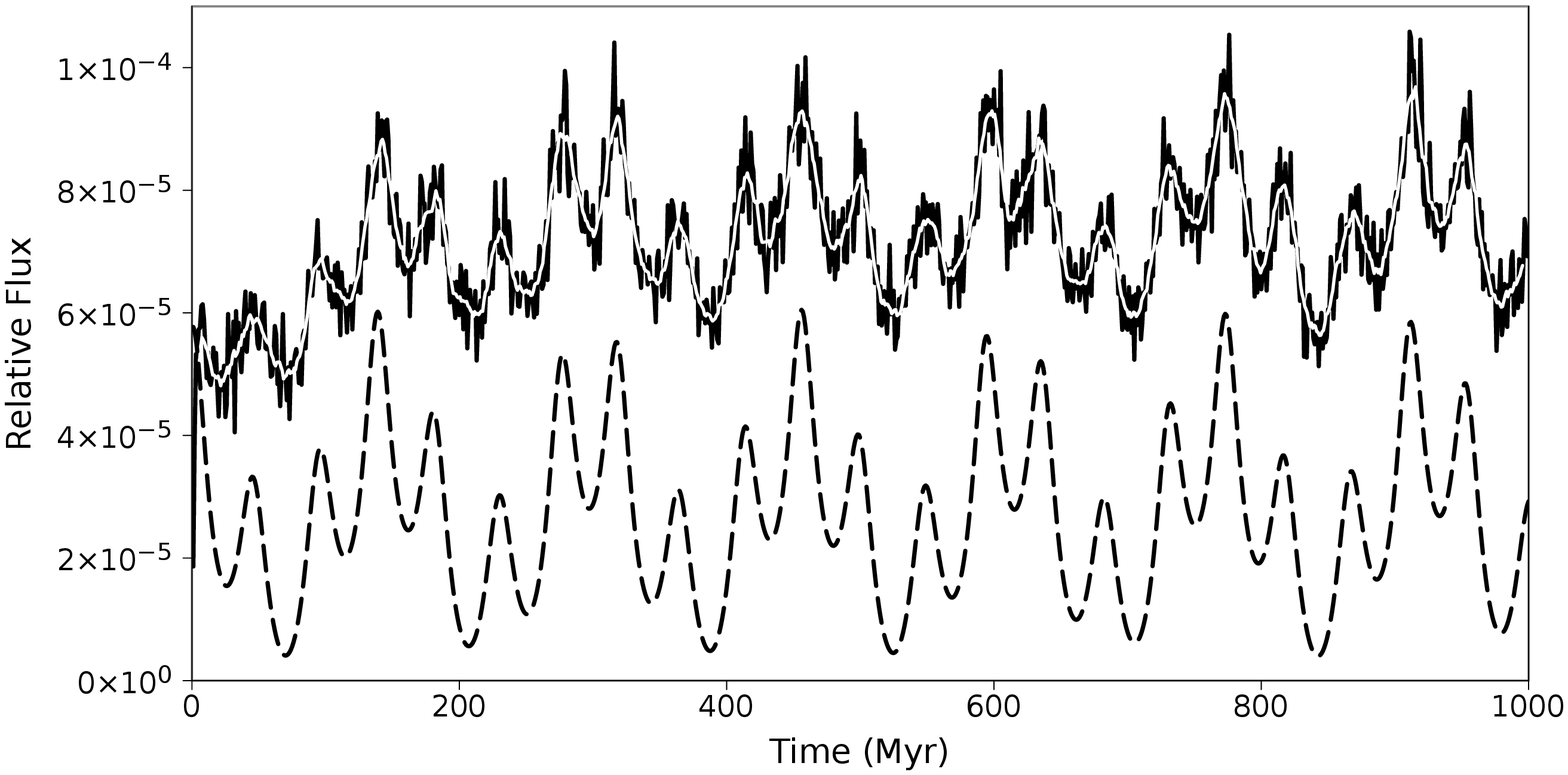}\\
  \includegraphics[width=84mm] {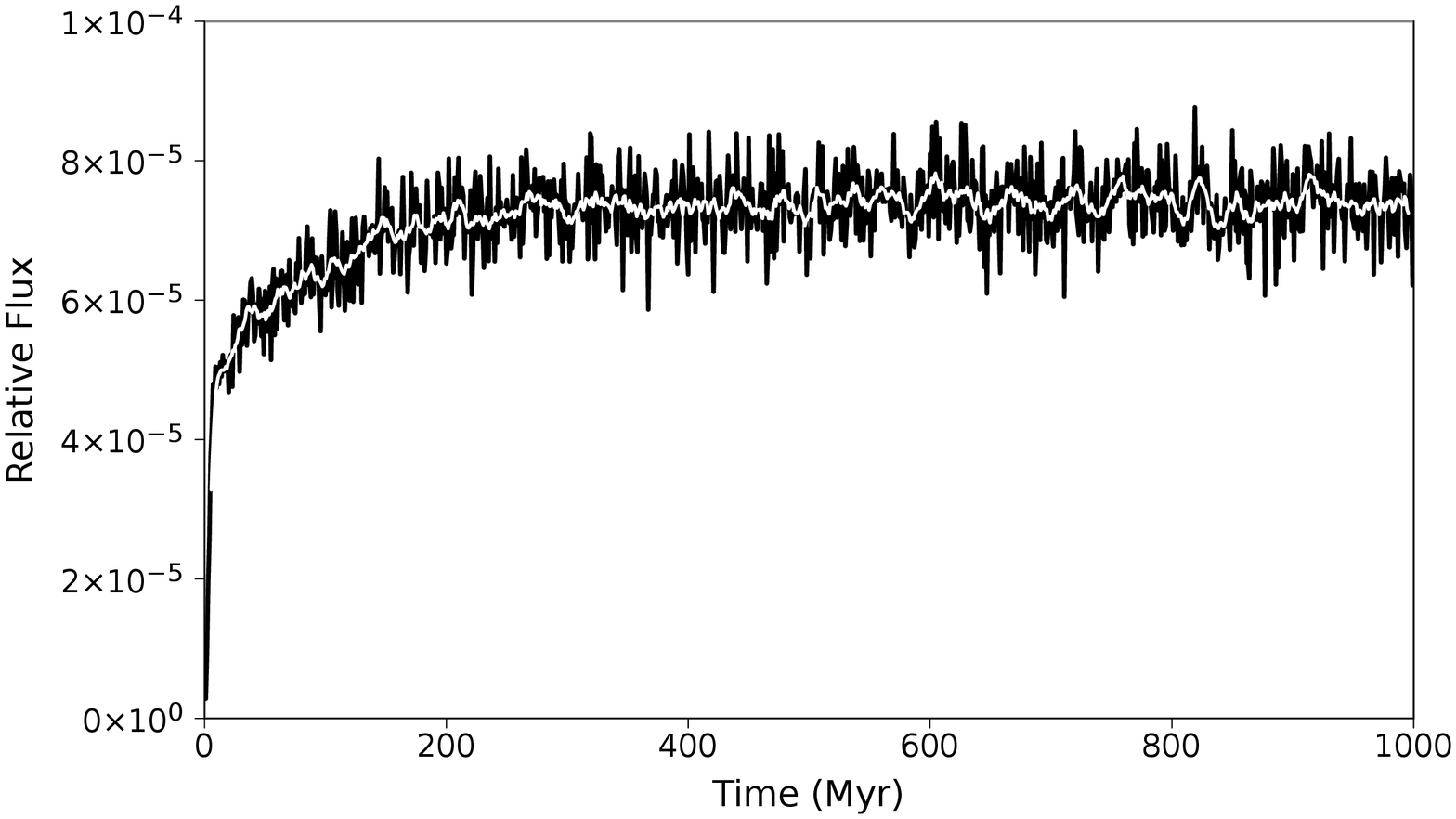} \caption{The
  time evolution of the relative flux of comets in the dynamic case (top
  panel), and the constant case (bottom panel). The solid line
  represents the relative flux of comets in 1 Myr bins, the white line a 10 Myr
  moving average. The dashed line in the top panel represents the
  evolution of the $G_3$-parameter.}\label{ts-dynamic}
\end{figure}

There is a clear difference between the two models when we look at the
temporal evolution of the cometary flux. The top panel of Fig.
\ref{ts-dynamic} shows, for the 'dynamic' case (and after a relaxation
time of about 100 Myr) that the mean cometary flux (shown as a 10 Myr
moving average) is well correlated with the changes in $G_3$ (dashed
line). The resulting fluxes from the simulation have been weighted
according to their relative number-density ($a^{-1.5}$), the resulting
total amount of comets in the simulation, by using this weighting
system is 1.8 $\times 10^6$. From Section \ref{motion-section}, there
are two main causes for the evolution of the flux, the major being the
radial motion, the minor being the vertical motion. The top panel of
Fig \ref{ts-dynamic} shows the major trend clearly following the
radial motion. The vertical motion is also followed, as is clear in
the 10 Myr moving average (which smooths out some of the Poisson noise
in the individual 1 Myr samples). We also ran a separate simulation
with a circular orbit, and with the vertical component intact, the
resulting fluxes followed perfectly the vertical motion, as was
expected due to the changes in $G_3$ being solely contributed by the
vertical component. In the case where the values for $G_i$ have been
kept constant (i.e. corresponding to a completely circular orbit with
no vertical motion), there is no evidence of any evolution, as seen in
the bottom panel of Fig. \ref{ts-dynamic}.

Simple Fourier-analysis of the dynamic flux (in the top panel of Fig.
\ref{ts-dynamic}) finds two distinct periods in the cometary flux. The
strongest signal is produced by a period of 143$-$167 Myr, and an
equal signal from a period of 41$-$45 Myr. The former corresponds to
the radial period (152 Myr) and the latter to the half-period of the
vertical period (43 Myr). The vertical signal is found to lie in the
range 41$-$45 Myr, and is quasi-periodic, as has been
seen earlier by \cite{Matese2001}.

\subsection{Comparison with earlier results}\label{flux-section}

\cite{Matese1995} derived various periods for the vertical Solar
oscillation, depending on the adopted model of the disc.  For example,
their `No-Dark-Disk Model' has a crossing-period very close to ours
(43 Myr), while their `Best-fit Model' produces a much lower period of
33 Myr. They assumed a $W$-velocity of 7.5 km s$^{-1}$, compared to
ours of 7.25 km s$^{-1}$, and does not cause much qualitative
difference in the vertical period of the orbit of the Sun. The lower
period comes from assuming a considerably higher mid-plane density
($\rho \approx 0.13$ M$_\odot$/pc$^3$) than ours.


\cite{Matese2001} found that the radial period of Solar motion modulates the
vertical period. However, they used very high local densities of matter in the
disc, so that the vertical oscillations dominated the radial oscillations.
This meant that the flux of comets into the inner Solar System with each
passage through the disc was greatly amplified compared to our simulations. We
find that the cometary flux varies with an amplitude of about 20\%, whereas
\cite{Matese2001} find flux variations of about a factor of two. The much
smaller amplitude in the signal which we advocate, even taking into account the
radial oscillations, would make finding a period in the scant cratering record
very difficult.

\cite{Fouchard06} calculated that a local mass density of $\rho = 0.1$
M$_\odot/$pc$^3$ corresponds to a cometary flux of around 10$^4$ for
their first 500 Myr interval for a source population of 10$^6$ comets,
where observed comets have $q \le$ 15 AU. Assuming that there are
10$^{12}$ comets in the Oort cloud. This produces a flux of $\sim$20
comets/yr, which is a factor of two less than our estimate.

The Galactic tide case in \cite{Rickman08} gives a flux of 100 comets
per 50 Myr, from a source population of 10$^6$ comets, where observed comets
have evolved from $q >15 AU$ to $q <$ 5 AU. This corresponds to a flux
of 2 comets/yr, again assuming an Oort cloud with 10$^{12}$ comets. Using
similar analysis we find that we get a flux of 290 comets per 50 Myr,
corresponding to a factor of three larger than \cite{Rickman08}.


A review by \cite{Bailer-Jones09} correlates terrestrial events with
cometary signals. One of the more interesting ones in the review is
the 140 $\pm$ 15 Myr period found by \cite{Rohde05} in the number of
known marine animal genera as a function of time. While this could
correspond to the Sun's radial period, \cite{Bailer-Jones09} considers
that it has not been significantly detected, the main problem being
that the entire time-span of the data covers no more than three
oscillations. Many other periods, proxies, and studies are mentioned
in \cite{Bailer-Jones09}, although the conclusion is that there is no
proven impact on biodiversity as a result of the orbital motion of the
Sun. 

\section{Discussion and conclusions} 

We have studied the long-term dynamics of Oort cloud comets under the influence
of both the radial and the vertical components of the Galactic tidal
field. Other perturbing forces on the comets, such as passing stars or passage
through spiral arms are ignored, since we aim to study the influence of just
the axisymmetric Galactic tidal field on the cometary motion.

We use an axisymmetric model of the Galaxy, and a recently revised value for
the Solar motion by \cite{Schonrich10}. This leads to vertical oscillations of
the Sun with an amplitude of about 100 pc, and radial oscillations over
about 1 kpc. The changing tidal forces on the Oort cloud are computed as the
Sun orbits for 1 Gyr in this potential, and the flux of comets entering the
inner Solar System is computed in simulations. 

As expected, the cometary flux is strongly coupled to the $G_3$-parameter in
the tidal forces, which is dominated by the local mass density seen along the
Solar orbit. Both the radial and vertical motions of the Sun can be seen in the
cometary flux, although the amplitude of the variations is small, implying that
detecting such a signal from the small number of age-dated craters would be
very difficult. This agrees with the recent review of the detectability of
the Solar motion in terrestrial proxies \citep{Bailer-Jones09}.

As $G_3$ is directly coupled to local density, it is easily affected
by the non-axisymmetric components in the motion of the Sun in the
Galaxy. This implies that spiral arms, a Galactic bar, giant molecular
clouds, or any other intermittently encountered structure should have
an effect on the flux. 

\section*{Acknowledgements}
PN wants to thank the Academy of Finland for the financial support in this
work. EG acknowledges the support of the Finnish Graduate School in Astronomy
and Space Physics.

\bibliographystyle{mn2e}
\bibliography{Oort_cloud_dynamics_in_MW}

\label{lastpage}

\end{document}